# Relevant Feature Selection Model Using Data Mining for Intrusion Detection System

Ayman I. Madbouly[#1], Amr M. Gody[#2], Tamer M. Barakat[#3]

[1]*Department of Research and Consultancy, Deanship of Admission and Registration, King Abdulaziz University*

*Jeddah, Saudi Arabia*

[2, 3]*Electrical Engineering Department, Faculty of Engineering, Fayoum University*

*Fayoum, Egypt*

*Abstract*— Network intrusions have become a significant threat in recent years as a result of the increased demand of computer networks for critical systems. Intrusion detection system (IDS) has been widely deployed as a defense measure for computer networks. Features extracted from network traffic can be used as sign to detect anomalies. However with the huge amount of network traffic, collected data contains irrelevant and redundant features that affect the detection rate of the IDS, consumes high amount of system resources, and slowdown the training and testing process of the IDS. In this paper, a new feature selection model is proposed; this model can effectively select the most relevant features for intrusion detection. Our goal is to build a lightweight intrusion detection system by using a reduced features set. Deleting irrelevant and redundant features helps to build a faster training and testing process, to have less resource consumption as well as to maintain high detection rates. The effectiveness and the feasibility of our feature selection model were verified by several experiments on KDD intrusion detection dataset. The experimental results strongly showed that our model is not only able to yield high detection rates but also to speed up the detection process.

## I. INTRODUCTION

With the increment use of networked computers for critical systems and the widespread use of large and distributed computer networks, computer network security attention increases and network intrusions have become a significant threat in recent years. Intrusion detection system (IDS) has been widely deployed to be a second line of defense for computer network systems along with other network security techniques such as firewall and access control. The main goal of intrusion detection system is to detect unauthorized use, misuse and abuse of computer systems by both system insiders and external intruders. There are various approaches to build intrusion detection systems. Intrusion detection systems can be classified into two categories based on the technique used to detect intrusions: anomaly detection and misuse detection [1], [2], [3]. Anomaly detection approach establishes the profiles of normal activities of users, operating systems, system resources, network traffic and services using the audit trails generated by a host operating system or a network scanning program. This approach detects intrusions by identifying significant deviations from the normal behavior patterns of these profiles. Anomaly detection approach strength is that prior knowledge of the security breaches of the target systems is not required. So, it is able to detect not only known intrusions but also unknown intrusions. In addition, this approach can detect the intrusions that are achieved by the abuse of legitimate users or masquerades without breaking security policy [4], [5]. The disadvantages of this approach were it had high false positive detection error, the difficulty of handling gradual misbehavior, and expensive computation [4], [6], [7]. On the other hand, misuse detection approach defines suspicious misuse signatures based on known system vulnerabilities and a security policy. Misuse approach probes whether signatures of known attacks are present or not in the auditing trails, and any matched activity is considered an attack. Misuse detection approach detects only previously known intrusion signatures. The advantage of this approach is that it rarely fails to detect previously notified intrusions, i.e. lower false positive rate [5], [8]. The disadvantages of this approach are it cannot detect new intrusions that have never previously been detected, i.e. higher false negative rate. Also, this approach has other drawbacks such as the inflexibility of misuse signature rules and the difficulty of creating and updating intrusion signature rules [4], [8], [9]. Intrusion detection system can also be classified based on the source of information collected into two types: host-based and network-based. Host-based systems analyze data collected from the operating system or applications running on the host subject to attack. Network-based systems look for sign of intrusions from network traffic being monitored. Most of commercial IDSs available use only misuse detection because most developed anomaly detectors still cannot overcome the limitations described before. This trend motivates many research efforts to build effective anomaly detectors for the purpose of intrusion detection. Researchers applied many anomaly detection techniques to intrusion detection. Vast majority of these researches concentrated on mining various types of data collected from raw network traffic or system audit data in order to build more accurate IDS [10], [11], [12]; that correctly classify alarms into attack and benign categories [13], [14], [15]. In this paper, anomaly intrusion detection system that detects anomalies by observing network traffic was considered. Features extracted from network traffic can be used to detect such anomalies, with high traffic and large scale networks there is a large amount of features' data to be observed for detection. To improve the detection accuracy and optimize the computational time and resources usage, intrusion detection system need to collect and study only the most relevance features that best distinguish between normal and attack traffic. An efficient data mining and machine learning techniques should be implemented on existing intrusion detection dataset to select best relevance features subset which provides the best accuracy and removes distractions [16]–[20].

The rest of the paper is organized as follows: Section II presents some related researches on intrusion detection which cover the feature selection and data mining. Section III briefly describes the KDD dataset used in this research. Section IV explains the details of the dataset pre-processing phase of the proposed model. The proposed model is presented in Section V. Finally, the experimental results and analysis are presented in Section 6 followed by some conclusions in the final section.

## II. RELATED WORK

Anomaly detection approaches have been extensively developed and researched. Many algorithms and methodologies have been suggested for anomaly detection [21]–[27]. These include machine learning [28]–[30], data mining [31]–[35], statistical [36], [37], neural networks [38], [39], information flow analysis [40], and approaches inspired from human immunology [26], [41], [42]. Recently, Data mining and machine learning have been widely used to solve many IDS classification problems. Many effective classification algorithms and mining techniques have been employed including traditional classification [43], [44], and hybrid classification [45]–[49].

Despite the existence of such different algorithms and approaches, it was found that none of them is able to detect all types of intrusion attacks efficiently in terms of the detection accuracy and classifier performance. As a result, recent researches aim to combine the hybrid classification strategy and features selection approaches using data mining to enhance the detection accuracy of models built for intrusion detection and to make smart decisions in determining intrusions. Siraj et al. [50] presented a hybrid intelligent approach for automated alert clustering and filtering in intrusion alert analysis in terms of classification accuracy and processing time. Panda et al. [48] proposed a hybrid intelligent approach using combination of data filtering along with a classifier to make intelligent decision that enhance the overall IDS performance. The proposed approach used combining classifier strategy in order to make intelligent decisions, in which data filtering is done after adding supervised classification or unsupervised clustering to the training dataset. Then the filtered data is applied to the final classifier methods to obtain the final decision. Agarwal et al. [47] proposed hybrid approach for anomaly intrusion detection system based on combination of both entropy of important network features and support vector machine (SVM). They evaluated the proposed approach using DARPA Intrusion Detection Evaluation dataset. They showed that hybrid approach outperforms entropy based and SVM based techniques. They identified that the fixed threshold range for entropy is the problem with the entropy based model, i.e. this method is not dynamic with the changes in normal conditions causing high false alarm rates. Also, they concluded that SVM based model alone does not give very good results as network features are used for learning without processing. To overcome these problems, they proposed the hybrid approach in which they calculated the normalized entropy of network features and sent it to SVM model for learning the behavior of the network traffic, and then this trained SVM model can be used to classify network traffic to detect anomalies. In their study [51], Mukherjee et al. showed the importance of reducing features in building effective and computationally efficient intrusion detection system. They proposed feature vitality based reduction method (FVBRM) to identify a reduced set of important input features using NSL-KDD dataset. They compared this method with three different feature selection algorithms, involving Gain Ratio (GR), Correlation-based Feature Selection (CFS), and Information Gain (IG). Feature reduction was performed on 41 NSL-KDD features and results obtained showed that (**10, 14, 20** and **24**) were the best features that are selected by CFS, GR, IG and FVBRM respectively. Their proposed method has improved classification accuracy compared to other feature selection methods but takes more time. Alhaddad, Mohammed J., et al. [52] carried out an experiment to study the applicability of different classification methods and the effect of using ensemble classifiers on the classification performance and accuracy. They compared the use of Naïve Bayes, and decision trees j48 and Random Forest classifiers as a single classifier. Also, they compared the use of AdaBoost.M1, and Bagging as ensemble classifier with both Naïve Bayes and J48 classifiers. Results showed that decision trees ensembles perform better than Naïve Bayes ensembles. Also, they conclude that Random Forest and Bagging perform better compared to AdaBoost.M1. They also observed that the performance of single decision tree is quite comparable with the decision trees ensembles, so it can be used if the performance requirements are not very strict and an optimized solution is needed for balancing between both system requirements and performance. Kumar et al. [53] proposed a new collaborating filtering technique for pre-processing the probe type of attacks and implemented a hybrid classifiers based on binary particle swarm optimization (BPSO) that has a strong global search capability and random forests (RF) a highly accurate classifier algorithm for the classification of PROBE attacks in a network. In their research they used PSO for fine-tuning of the features whereas RF is used for Probe type of attacks classification. Recently, Swarm intelligence was used in many researches to optimize the features selection process, in their study, Elngar, Ahmed A. et al. [54], proposed a hybrid intrusion detection system where particle swarm optimization (PSO) is used as a feature selection algorithm and C4.5 decision tree (DT) as a classifier. They evaluated their proposed system by several experiments on NSL-KDD benchmark network intrusion detection dataset. Results obtained showed that the proposed approach could effectively increase the detection accuracy and minimize the timing speed. By reducing the number of features from 41 to 11, the detection performance was increased to 99.17% and the time was speeded up to 11.65 sec. They also compared PSO feature selection method with a well-known method Genetic Algorithm (GA), results showed that the proposed PSO-DT intrusion detection system gives better detection performance than GA-DT and also less model building time. In another study, Chung et al. [55] proposed a new hybrid intrusion detection system by using intelligent dynamic swarm based rough set (IDS-RS) for feature selection and simplified swarm optimization for intrusion data classification. IDS-RS is proposed to select the most relevant features that can represent the pattern of the network traffic. In order to improve the performance of SSO classifier, a new weighted local search (WLS) strategy incorporated in SSO is proposed. The purpose of this new local search strategy is to discover the better solution from the neighbourhood of the current solution produced by SSO. The performance of the proposed hybrid system on KDD Cup 99 dataset has been evaluated by comparing it with the standard particle swarm optimization (PSO) and two other most popular benchmark classifiers. The testing results showed that the proposed hybrid system can achieve higher classification accuracy than others with 93.3% and it can be one of the competitive classifier for the intrusion detection system. Suthaharan et al. [56], suggested an approach that analyzed the intrusion datasets to evaluate the most relevant features to a specific attack, and determined the level of contribution of each feature and eliminate it from the dataset automatically. They adopted the Rough Set Theory (RST) based approach and selected relevance features using multidimensional scatter-plot automatically. Gradually feature removal method was proposed by Yinhui Li et al. [57] to choose the critical features that represent various network attacks, 19 features were chosen. With the combination of clustering method, ant colony algorithm and support vector machine (SVM), they developed an efficient and reliable classifier to judge a network visit to be normal or not with an accuracy of 98.6249%. Yang Li et al. [58] proposed a new wrapper-based feature selection algorithm using search strategy based on modified random mutation hill climbing (RMHC) with linear support vector machine (SVM) as evaluation criterion to obtain the optimum features subset from KDD dataset features. Results showed that the proposed approach could be used to build lightweight IDS with high detection rates and low computation cost. Olusola et al. [59] discussed the selection of relevance of each feature in the KDD-99 where rough set degree of dependency and dependency ratio of each attack type were employed to determine the most discriminating features for each attack type. Results showed that two features (**20**, **21**) are not relevant to any attack type, i.e. have no relevance in intrusion detection process. Also, results showed that features (**13**, **15**, **17**, **22**, **40**) are of little significant in the intrusion detection.

Naming significant and important features relevant to specific attack class have been discussed in [60]–[63]. In their study, S. Zargari, and D. Voorhis [64] attempted to explore significant features in intrusion detection in order to be applied in data mining techniques. They examined the effect of samples sizes on the performance of intrusion detection by selecting five random samples with various sizes with the same distribution of the attacks in the samples as the distribution of the attacks in the KDD-dataset. They used a voting system to propose a subset of 4-features (**3**, **5**, **6**, **39**) based on results summarized from five previously published papers that named the significant features relevant to the four types of attacks in the KDD. This proposed subset was compared with other subset selected using different attribute selection evaluators and methods using Weka. The conclusions drawn from comparing the results showed that the proposed four features improved the detection rates better than the other four features suggested by Weka methods when only a subset of four features are considered in the data mining, and can be used in instead of 41 features in the KDD-dataset to reduce the dimensionality of the dataset while the detection rate is not much affected.. Also they concluded that a higher rate of detection is achieved if the information gain method with ten features (**5**, **3**, **23**, **24** + **33**, **35**, **36**, **34**, and **6**) is used.

### III. DATASET DESCRIPTION

Intrusion detection system software is used to detect network intrusions and protect computer network from unauthorized users, including perhaps insiders. The task of intrusion detector learning is to build a classifier (predictive model) that can distinguish between "bad" connections, called intrusions or attacks, and "good" normal connections. At 1998 MIT Lincoln Labs prepared and managed the 1998 DARPA program, an Intrusion Detection Evaluation Program. The objective of DARPA was to survey and evaluate research in intrusion detection. A data set of a wide variety of intrusions simulated in a military network environment was provided as standard audited set of intrusion data [65]. Nine weeks of raw TCP dump data was acquired from Lincoln Labs setup of an environment for a local-area network (LAN) that simulate a typical U.S. Air Force LAN, Fig. 1. In DARPA program the simulation LAN was operated as if it were a true Air Force environment LAN, but peppered it with multiple attacks. The Knowledge Discovery and Data Mining (KDD'99) intrusion detection contest uses a version of this DARPA dataset. KDD'99 [65] was developed, by the Massachusetts Institute of Technology - MIT, during the international competition on data mining in 1999. KDD'99 is one of the most popular benchmark datasets used to choose proper intrusion detection metrics.

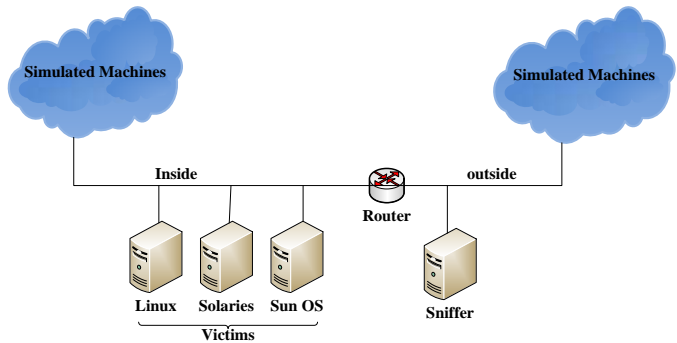

Fig. 1 DARPA LAN setup, simulating a typical U.S. Air Force LAN [65]

In KDD'99 dataset each TCP Connection is represented by 41 Features. Here, TCP connection refers to a sequence of TCP packets transmitted over a well-defined protocol from a source IP address to a target IP address well-defined starting and ending times. About five million TCP connection records of raw training data (four gigabytes of compressed binary TCP dump) were processed from TCP dump data during seven weeks of network traffic. Similarly, around two million connection records of test data were processed from during two weeks. Each TCP connection was labeled as "normal" or "attack" with a specific attack type; the length of each connection record is 100 bytes. Simulated attack types fall in one of the following four categories [66]:

A. *Denial of Service Attack (DOS):*

Is an attack in which the attacker makes some computing or memory resource too busy or too full to handle legitimate requests, or denies legitimate users access to a machine.

B. *User to Root Attack (U2R):*

Is a class of exploit in which the attacker starts out with access to a normal user account on the system (perhaps gained by sniffing passwords, a dictionary attack, or social engineering) and is able to exploit some vulnerability to gain root access to the system.

C. *Remote to Local Attack (R2L):*

Occurs when an attacker who has the ability to send packets to a machine over a network but who does not have an account on that machine exploits some vulnerability to gain local access as a user of that machine.

D. *PROBE Attack:*

It's any attempt to gather information about a network of computers for the apparent purpose of circumventing its security controls.

Attacks could be classified based on the combination of these 41features. To understand how each of these features contributes to this classification problem, features were gathered into four groups [67], [66], as follows:

*Basic features*: features that identify packet header properties which represent connection critical metrics.
*Content features*: features represent useful information extracted from the packets that help experts to identify known forms of attacks.
*Time based traffic features*: features that are computed with respect to a 2-seconds time interval window. These could be divided into two groups, same host features and same service features.
*Host based traffic features*: features that are computed with respect to a connection window of 100 connections. Statistics are calculated from a historical data that is estimated from the last hundred used connections to the same destination address. These are useful to detect slow probing attacks that scan hosts or ports using at much larger time interval than 2 seconds.

IV. DATASET PRE-PROCESSING

KDD'99 is actually composed of three datasets:

1) *'Whole KDD':* this is the original dataset created out of the data collected by using the implemented sniffer, and it's the largest one, it contains about 4 million registers.
2) *'10% KDD':* to reduce the computational cost of the processed data the amount of data to be processed need to be reduced as must as possible. Thus this subset contains only 10% training data that is randomly taken from the original dataset. This is usually used to train IDS systems.
3) *'Corrected KDD':* this is a testing dataset that has different distribution of probability of attacks compared to the other two data set. This data set includes specific attack types that are not included the training data attack types. It includes 14 new types of attacks and usually used to test IDS systems. These new types of attacks included in the test dataset allow this dataset to be more realistic while using it to evaluate intrusion detection systems.

The complete dataset contains a total of 24 attack types inside the training dataset, and an additional 14 attack types in the test dataset. As mentioned by [66], [68], [69], KDD dataset has some problems that cause the evaluation results unreliable. One important problem is the large number of redundant records that bias learning algorithm to the classes with large repeated records. While less repeated records such as U2R and R2L usually more harmful to network will not be learned. Also, more repeated records results in biased evaluation results.

To solve this issue, all repeated records in the '10% KDD' train dataset and 'Corrected KDD' test set were deleted, and kept only non-redundant records. Table 1 and Table 2 show the class distribution and statistics of the reduction of repeated records in the KDD train and test datasets, respectively.

Out of the reduced training dataset four class-based datasets have been constructed: DOS, PROBE, R2L, U2R each of these four datasets contains the attack type records + the NORMAL class records. These four datasets were used along with the ALL classes dataset to find the best relevant features as will be explained later. During our early experiments it was noticed that most of the classification error of the PROBE class happened because of these instances that were misclassified as DOS class type and of course as NORMAL class type. For this another dataset that contains only DOS+PROBE classes was constructed. This will be used to find the best features that can be used to distinguish the two classes.

TABLE I
10 % KDD TRAINING DATASET CLASS DISTRIBUTION

| Class | Number of Instances | | | |
|---|---|---|---|---|
| | Before Removing Duplicate Instances | After Removing Duplicate Instances | % of Reduction | % To Total # of Instances |
| Normal | 97278 | 87832 | 9.7% | 60.33% |
| DOS | 391458 | 54572 | 86.1% | 37.48% |
| PROBE | 4107 | 2131 | 48.1% | 1.46% |
| R2L | 1124 | 997 | 11.3% | 0.68% |
| U2R | 54 | 54 | 0.0% | 0.04% |
| Total Number of Instances | 494021 | 145586 | 70.5% | |

TABLE II
CORRECTED KDD TEST DATASET CLASS DISTRIBUTION

| Class | Number of Instances | | | |
|---|---|---|---|---|
| | Before Removing Duplicate Instances | After Removing Duplicate Instances | % of Reduction | % To Total # of Instances |
| Normal | 60593 | 47913 | 20.9% | 61.99% |
| DOS | 229855 | 23570 | 89.7% | 30.50% |
| PROBE | 4166 | 2682 | 35.6% | 3.95% |
| R2L | 16345 | 3056 | 81.3% | 3.47% |
| U2R | 70 | 70 | 0.0% | 0.09% |
| Total Number of Instances | 77291 | 311029 | 75.1% | |

V. PROPOSED MODEL

A model that consists of four stages has been proposed as shown in Fig. 3. *The first stage*, **Data Pre-Processing**, is the stage in which data was prepared as discussed in the previous section.

*The second stage*, **Best Classifier Selection**, is the stage in which the classifier that has the best performance and accuracy has been chosen to be our model classifier. Actually, machine learning has been attempted by many researchers to innovate complex learners that optimize detection accuracy over KDD'99 dataset [66]. Out of these machine learning techniques, there are many widely used. For example, Naive Bayes (NB) [70], Random Forest Tree (RFT) [71], Random Tree (RT) [72], J48 decision tree learning [73], NBTree [74], Multilayer Perceptron (MLP) [75], and Support Vector Machine (SVM) [76]. However, researches showed that using meta-classifiers/ensemble selection give better performance [77].

In our experiment, performance and accuracy results of four learning techniques recently mentioned by many researches as best classifiers approaches were compared. Our ensemble classifier consists of a boosting algorithm, Adaboost M1 method [78], with four different learning techniques: Naive Base, Random Forest, SVM and J48. Learning and classification were done using the KDD training dataset resulted from the previous stage with the 41 features exist. The best classifier selected in this stage was used in the next stages to evaluate the selected features set and to measure the detection accuracy and performance using this selected best set.

*The third stage*, **Feature Reduction**, was implemented to reduce the **41** features set by deleting redundant features that have no importance in the detection process. Two ranked features lists have been deduced, one for features that are mostly selected by different algorithms, as shown in Table 4. And, the other one for features that are most important to all attack classes, as shown in Table 5. Common features that came at the end of these two ranked lists are then excluded one by one. These less important features were deleted one by one and after deleting each feature the rest of features were used to re-evaluate the detection system again to make sure that deleting these features did not affect the overall detection accuracy and performance.

*The fourth stage*, **Best Features Selection**, consists of two separate phases; **Gradually ADD Feature** and **Gradually DELETE Feature**. The idea here is to use two different techniques to select the best features.

In the first phase, Gradually ADD Feature, the smallest features set that is commonly selected by all attack types and were selected by all algorithms with high rates were chosen to be the start set of this phase. Detection accuracy and performance were calculated using this smallest feature set, after that one feature was added each time to the set and the detection system was re-evaluated. If the detection and accuracy is increased the start features set was updated by adding this feature to the best features set. Since U2R is the attack type with the lowest number of instances in our dataset, features that are important to detect U2R attack type are firstly used. Each feature in the U2R's important features set was added to the smallest common set and the feature that achieves the best detection performance is added to the set. The same process is repeated for the features that are important to detect R2L attack type and the PROBE attack type in order. A feature was added to the best features set if it increases (or at least does not decrease) the detection performance of all attack types not only of detection performance of that attack type this feature is important for.

In the second phase, Gradually DELETE Feature, the reduced features set deduced from third stage was used as the start features set. By ranking this features set, and delete one of the low importance feature each time from the bottom of the ranked list then re-evaluate the detection performance and accuracy, it was decide to keep or to remove that feature from the features set. Deletion was started by deleting the lowest feature from the DOS most important features set. Same Process is repeated for PROBE, R2L and U2R important feature set in order.

Comparing results from both phases, it was concluded that the best features set that achieved an optimized detection performance and accuracy compared to the full 41 features set was the following 11 features set {**1, 3, 4, 5, 6, 10, 14, 23, 25, 30, 35**}.

To implement the last two stages, feature selection process was conducted. Feature selection process involves searching through all possible combinations of features in our dataset to find the best subset of features works best for attack prediction and classification process.

WEKA 3.7.7 machine learning tool [79] was used to calculate the best feature subset. In Weka, it's needed to setup two objects, an *attribute evaluator* and a *search method*. The attribute evaluator determines what method is used to assign a worth to each subset of features, while the search method determines what style of search is used to search through the space of feature subsets. In our experiments, seven different search methods were used as shown in Table 4. This gives us a preliminary overview of the most important features out of the 41 features. To find the minimum subset of features that works best for attack detection and the classification process compared to the full 41 feature subset, Correlation-based Feature Subset Selection (CFS) was used as the attribute evaluator with the seven search method algorithms. Results are shown in Table 4. All experiments were performed using Windows® 7- 32 bit operating system platform with core i7 processor 2.4 GHz, 4.0 GB RAM.

## VI. EXPERIMENTAL RESULTS AND ANALYSIS

### A. Performance Measure

The detection effectiveness of the proposed detection system is measured in terms of *Specificity*, *Sensitivity or True Positive Rate (TPR)*, *FPR (False Positive Rate)*, *Negative Predictive Value (NPV)*, *Precision or Positive Predictive Value (PPV)*, *F-Measure*, *Matthews correlation coefficient (MMC)*, and the *Classification Accuracy* as defined by the following equations and equations shown in Table 3:

F-Measure = (2 * TP) / ((2 * TP) + FP + FN)     (1)

$$MMC = \frac{(TP*TN - FP*FN)}{\sqrt{(TP+FN)(TP+FP)(TN+FN)(TN+FP)}}$$     (2)

$$Accuracy = \frac{Correctly\ classified\ Instances}{Total\ number\ of\ Instances}$$     (3)

These measures are calculated based on the confusion matrix that shows the four possible binary classification outcomes as shown in Table 3.

Where:
**TN** (**T**rue **N**egatives): Indicates the number of *normal* events successfully labeled as *normal*.
**FP** (**F**alse **P**ositives): Refer to the number of *normal* events being predicted as *attacks*.
**FN** (**F**alse **N**egatives): The number of *attack* events incorrectly predicted as *normal*.
**TP** (**T**rue **P**ositives): The number of *attack* events correctly predicted as *attack*.

TABLE III
CONFUSION MATRIX

| Actual Class | Classified Class | | |
|---|---|---|---|
| | Normal | Attack | |
| Normal | TN | FP | Specificity ≡ [1-PR]= $\frac{TN}{TN+FP}$ |
| Attack | FN | TP | Sensitivity ≡ TPR= $\frac{TP}{TP+FN}$ |
| | NPV= $\frac{TN}{TN+FN}$ | PPV= $\frac{TP}{TP+FP}$ | |

TABLE IV
BEST FEATURES SELECTED BY ALL ALGORITHMS FOR EACH CLASS OF ATTACK

| Rank Search (Gain Ratio) | |
|---|---|
| All | 3,4,5,6,11,12,14,22,26,29,30,38,39,9,33,35,37,23,34 |
| DOS | 3,4,5,6,12,25,26,29,30,37,38,39 |
| PROBE | 25,27,29,37,17,30,4,5,26,38 |
| R2L | 5,10,11,18,22,26,9,39 |
| U2R | 14,17,1,18,29,39,9,11,13,32,33 |
| Best Features by Algorithm | 3,4,5,6,9,11,12,14,17,18,22,25,26,29,30,33,37,38,39 |

| Rank Search (Info Gain) | |
|---|---|
| All | 3,4,5,6,12,14,23,25,26,29,30,33,34,35,38,39,37,9,32 |
| DOS | 3,4,5,6,12,23,25,26,29,30,33,34,35,37,38,39 |
| PROBE | 3,4,5,6,12,23,25,27,29,30,33,34,35,37,40,38,36,41 |
| R2L | 3,5,6,10,22,33,36,9,26,16,37 |
| U2R | 3,14,17,18,29,39,11 |
| Best Features by Algorithm | 3,4,5,6,9,12,14,23,25,26,29,30,33,34,35,36,37,38,39 |

| Best Frist | |
|---|---|
| All | 3,4,8,10,12,25,29,30,37,9,32 |
| DOS | 4,5,12,29,30,37,26,6,25 |
| PROBE | 25,27,29,37,5,17,30,38 |
| R2L | 5,10,39,9,26,16 |

| | | |
|---|---|---|
| U2R | 1,14,17,18,29,39,11 | |
| **Best Features by Algorithm** | 4,5,9,10,12,17,25,26,29,30,37,39 | |
| **Evolutionary Search** | | |
| All | 3,4,12,25,29,30,37,6,38,8,10,11,5,31,39,9,14,15,16,18,32,33,36 | |
| DOS | 5,12,26,29,30,37,6,25,3,4,8,23,32,33,39,10,18,34,38,41 | |
| PROBE | 25,27,29,37,30,38,5,4,17,26,2,6,10,33,34,39,41 | |
| R2L | 10,5,26,39,9,16,22,11,36 | |
| U2R | 13,14,17,18,32,29,33,39,4,5,6,23,31,37 | |
| **Best Features by Algorithm** | 3,4,5,6,8,9,10,11,12,14,16,17,18,23,25,26,29,30,31,32,33,34,36,37,38,39,41 | |
| **Greedy Stepwise** | | |
| All | 5,6,12,22,25,26,29,30,35,37,39,23,4,9,17,31,18,14,11,38 | |
| DOS | 3,6,12,29,30,37,38,4,5,23,25,26,17,39 | |
| PROBE | 25,29,4,37,27,30,38,5,6 | |
| R2L | 10,11,22,6,9,26,16,5,39 | |
| U2R | 13,14,29,17,18,32,33,11 | |
| **Best Features by Algorithm** | 4,5,6,9,11,12,14,17,18,22,23,25,26,29,30,37,38,39 | |
| **PSO Search** | | |
| All | 3,4,8,10,12,25,29,30,37,9,32 | |
| DOS | 4,5,12,29,30,37,26,6,25,11 | |
| PROBE | 25,27,29,37,17,5,3 | |
| R2L | 5,10,39,9,26,16 | |
| U2R | 1,14,17,18,29,39,11 | |
| **Best Features by Algorithm** | 3,4,5,9,10,11,12,17,25,26,29,30,37,39 | |
| **Tabu Search** | | |
| All | 3,4,8,10,12,25,29,30,37,9,32 | |
| DOS | 5,12,29,30,37,26,25,6,11 | |
| PROBE | 25,27,29,37,17,5,3 | |
| R2L | 5,10,39,9,26,16 | |
| U2R | 1,14,17,18,29,39,11 | |
| **Best Features by Algorithm** | 3,5,9,10,11,12,17,25,26,29,30,37,39 | |

TABLE V
COMMON IMPORTANT FEATURES FOR EACH ATTACK CLASS

| | Common Important Features |
|---|---|
| All | 3,4,5,6,10,12,14,23,25,26,29,30,32,33,35,37,38,39 |
| DOS | 3,4,5,6,12,23,25,26,29,30,37,38,39 |
| PROBE | 3,4,5,6,17,25,27,29,30,37,38 |
| R2L | 5,9,10,16,22,26,39 |
| U2R | 1,13,14,17,18 |

*B. Results Analysis and Discussion*

In the proposed model, firstly our feature selection algorithm to select the best features set for all attack types (DOS, PROBE, R2L, U2R, and NORMAL class) was used. Then, a lightweight intrusion detection system using the selected best features set was built. All performance measures mentioned above are used to compare detection performance of our proposed system using the selected best features set with those using all 41 features set. Also, the performance of the proposed best features set in detecting known and new attacks compared to the all 41 features set were evaluated. The confusion matrix and time taken to build the classification model are also given for comparison.

The classification performance is measured by using ensemble classifier consists of a boosting algorithm, Adaboost M1 method [78], with Random Forest learning techniques. The classification was performed using a full training set and 10-fold cross-validation for the testing purposes. In 10-fold cross-validation, the available dataset are randomly divided into 10 separate groups of approximately equal size subsets. Nine subsets of these groups are used to build and train the classifier model and the 10[th] group is used to test the built classifier model and to estimate the classification accuracy. The whole operation is repeated 10 times so that each of the 10 subsets is used as test subsets once. The accuracy estimate of the 10-fold cross-validation method is the mean of the estimates for each of the 10 classification processes. 10-fold Cross-validation method has been used and tested extensively and has been found to work well in general when adequate and sufficient dataset is available [80], [81].

Fig. 2 shows a comparison between the class detection accuracy while using the best 11-features set and 41-features set. Results showed that the best 11-features set out perform the 41-features set in detecting U2R and R2L Attack types, the types with the lowest instances in the dataset. The best 11-features set have the same accuracy in detecting DOS attack type and almost same accuracy while classifying traffic as NORMAL. The most interesting result is the accuracy of detecting PROPE attacks. It seems that the best 11-features set have lower accuracy compared to the 41 features set, however out of the (**2131**) instances of PROBE attack in the dataset, using the best 11-features set the classifier could detect (**2076**) instances correctly, (**31**) instances was misclassified as DOS attack, (**1**) instance misclassified as U2R attack, and (**23**) instances were misclassified as NORMAL. However, using the 41-features set the classifier could detect (**2094**) instances correctly, (**10**) instances was misclassified as DOS attack, and (**27**) instances were misclassified as NORMAL, i.e. an addition of (**4**) attack instances are classified as NORMAL traffic while using the full 41-features set.

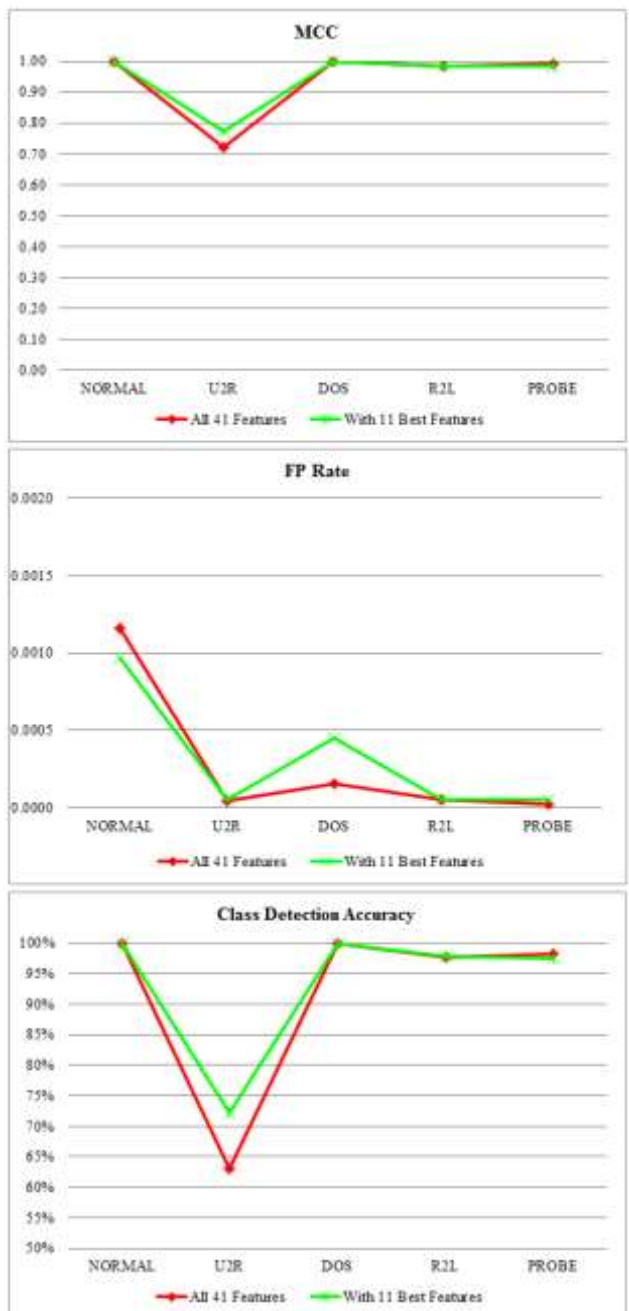

Fig. 2-a Comparison between different performance measures using the best 11-features set and 41-features set

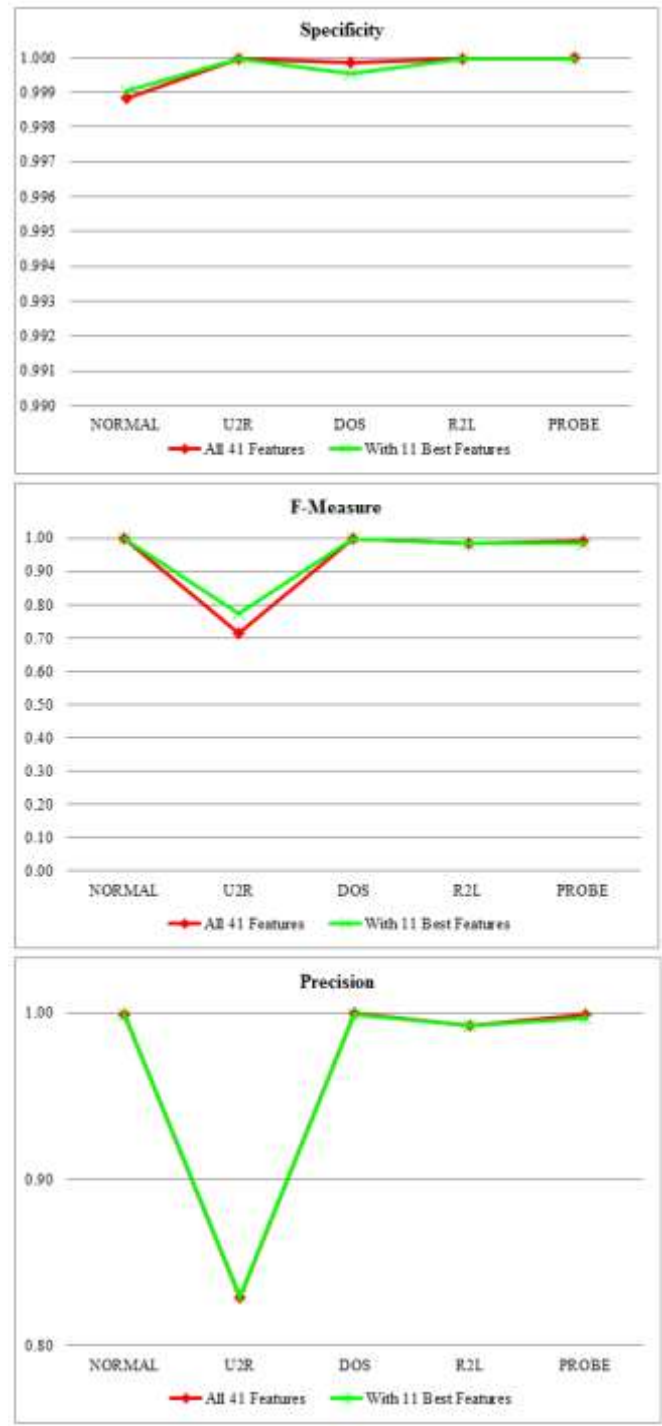

Fig. 2-b Comparison between different performance measures using the best 11-features set and 41-features set

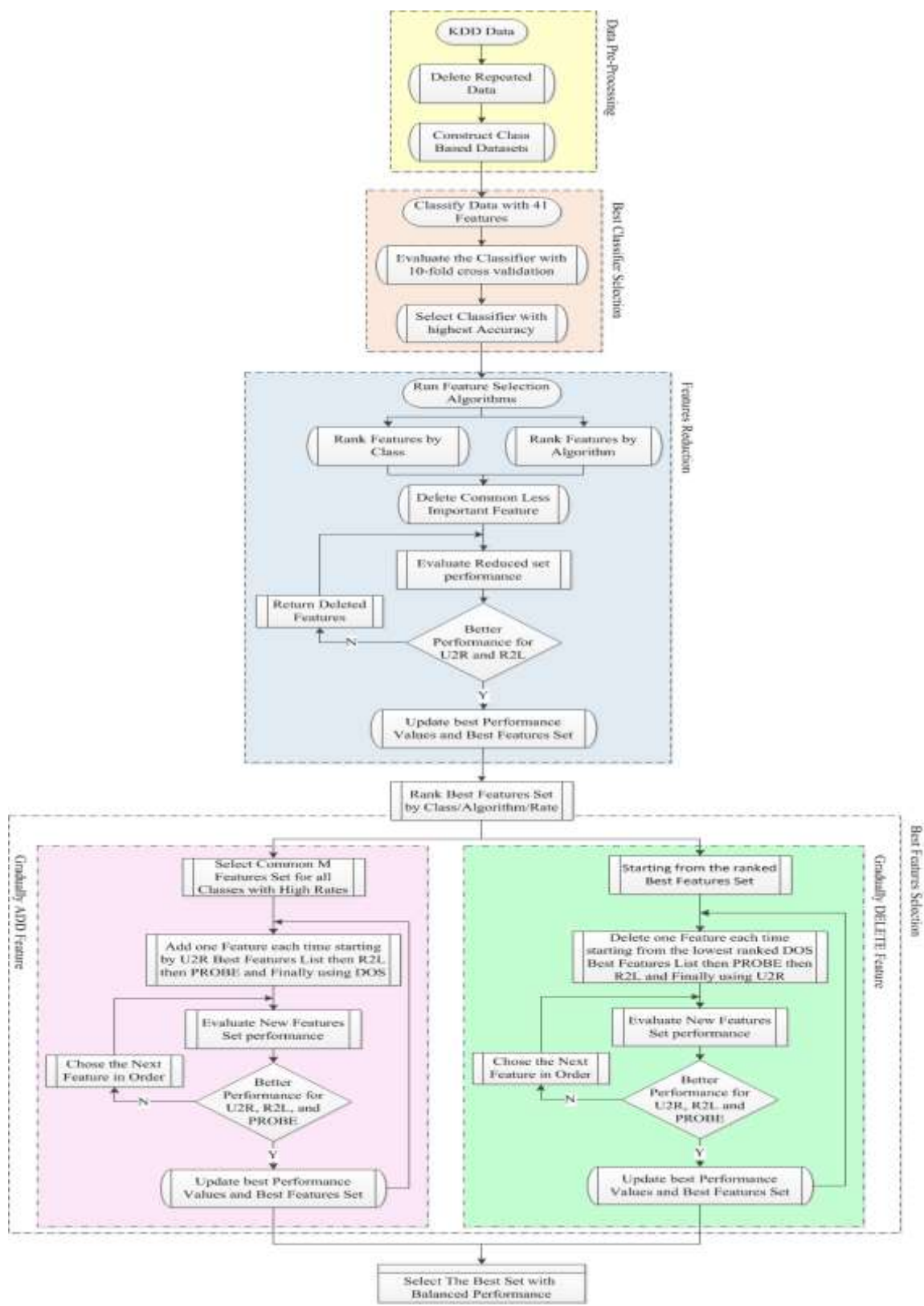

Fig. 3 The proposed model flowchart

VII. CONCLUSION

In this paper, relevant feature selection model was proposed to select the best features set that could be used to design a lightweight intrusion detection system. Feature relevance analysis is performed on KDD 99 training set, which is widely used by machine learning researchers. Seven different feature evaluation methods were used to select and rank relevant features. The proposed model has four different stages, *Data Pre-Processing, Best Classifier Selection*, *Feature Reduction*, and *Best Features Selection*. Redundant records existed in the train dataset that bias learning algorithm to the classes with large repeated records are deleted. Out of the reduced training dataset four class-based datasets have been constructed: **DOS, PROBE, R2L, U2R** each of these four datasets contains the *attack type records + the NORMAL class records*. Important features for each attack type have been determined. Our results indicate that certain features have no relevance or contribution to detect any intrusion attack type. Some features are important to detect all attack types, and certain features are important to detect certain attack types. A set of best **11**- features have been proposed and tested against the full **41**- features set. This reduced features set can be used to build a faster training and testing process, less resource consumption as well as maintaining high detection rates. The effectiveness and the feasibility of our feature selection model were verified by several experiments on KDD intrusion detection dataset. The experimental results strongly showed that our model is not only able to yield high detection rates but also to speed up the detection process.